\documentstyle[11pt,newpasp,twoside,epsf]{article}
\begin{document}
\title{Observing the CMB with the AMiBA}
\author{Ravi Subrahmanyan}
\affil{Australia Telescope National Facility, CSIRO,
	Locked bag 194, Narrabri, NSW 2390, Australia}

\begin{abstract}
I discuss the capabilities and limitations of the 
AMiBA for imaging CMB anisotropies.  Michael
Kesteven (ATNF-CSIRO) has proposed drift-scanning as an observing
strategy for measuring and rejecting any instrumental response
that the close-packed interferometers may have to the local environment.
The advantages of mosaic imaging CMB anisotropies using a co-mounted
interferometric array in a drift-scanning observing mode are discussed.
A particular case of mosaic imaging a sky strip using a two-element
AMiBA prototype interferometer is considered and the signal-to-noise
ratio in the measurement of sky anisotropy using this observing
strategy is analysed.
\end{abstract}

\section{Introduction}

The ASIAA, Taiwan, is building an Array for Microwave Background
Anisotropy (AMiBA) in collaboration with the National Taiwan
University.  The science goals are discussed by Lo et al. (2001).
The telescope would be an interferometric array of 19
elements co-mounted on a single fully steerable platform.  
The array may be configured using either elemental apertures
of 1.2 or 0.3~m diameters.  Dual polarization receivers would
have 20~GHz bandwidth centered at 95~GHz and the system temperature
is expected to be about 75~K.

\section{Imaging capabilities of the AMiBA}

Surveys with the AMiBA are proposed to be made with the 19 apertures
configured as a hexagonal close-packed array on the platform
to maximize the surface brightness sensitivity of the 
interferometers.  The configuration gives a well filled 
sampling of the u,v-domain with a radial density which peaks at 
the spacing corresponding to the distance between the centres of
the adjacent apertures (which, for engineering reasons, is somewhat
larger than the aperture diameters) and tapers off to zero at a 
u,v-distance five times larger.  When the visibility data from
single sky pointings are `naturally
weighted' --- for maximum flux sensitivity --- and images are synthesized, 
the synthetic beam has FWHM of 2.6~arcmin when the 1.2~m apertures
are used and 10.4~arcmin when the 0.3~m apertures are used.
The primary beam is expected to have HPW of 11~arcmin (1.2~m apertures)
and 44~arcmin (0.3~m apertures).
The absence of a `zero-spacing' measurement in the interferometer array
results in a `hole' in the u,v-coverage at the origin whose size depends
on the taper in the illumination at the edges of the apertures.  As a result,
the main lobe of the synthetic beam is surrounded by a negative bowl
which limits the surface brightness sensitivity.

Assuming aperture efficiencies of 0.6, we expect the 
1.2 and 0.3~m apertures to have sensitivities about 
4100 and 65100~Jy~K$^{-1}$ respectively; antenna temperatures from 
planets Jupiter and Saturn would be detectable with high
signal-to-noise ratio, for calibration, in seconds integration time.
When all four cross products are measured between the dual polarization
channels of every antenna pair, and images are synthesized in the
Stokes parameters using `naturally' weighted visibility data of all the 
171 baselines, the rms thermal noise after 1-hr integration time 
is expected to be 1.4 and 
22~mJy~beam$^{-1}$ respectively when 1.2 and 0.3~m apertures are used.

\section{Confusion limits}

I adopt the 90-GHz source counts derived by Holdaway, 
Owen \& Rupen (1994), with 180 sources all sky exceeding 1~Jy and
4400 sources all sky exceeding 100~mJy.  The rms confusion noise,
owing to these discrete synchrotron sources, is expected to be
less than 70 and 2~$\mu$Jy~beam$^{-1}$ in the images made respectively
with the 0.3 and 1.2~m aperture arrays.  The thermal noise will not
reach this confusion noise for any reasonable integration time: the images
will always be thermal noise limited.  Consequently, extragalactic synchrotron 
sources will only appear in AMiBA images above the instrument thermal noise 
as well separated discrete sources.  
The source counts imply that we may expect to
detect a source above 1-$\sigma$ thermal noise, 
in any AMiBA pointed observation,  only
after integration times of 50 and 300~hr respectively 
with the arrays made of 0.3 and 1.2-m apertures.  

Confusion from the primary CMB anisotropy may limit the SZE-cluster
survey sensitivity.  In Stokes I images made with the 171-baseline
array of 1.2-m apertures, the primary CMB contributes an rms image noise 
of about 2~mJy~beam$^{-1}$ which exceeds the thermal noise in 
1-hr integration.  However, the T-mode primary anisotropy is expected to
vanish at multipoles exceeding $l \sim 2500$ corresponding to 
a u,v-distance of 400: all of the T-mode power will be confined to the
1.2-m baselines and may be discriminated using appropriate filtering
in $l$-space.

\section{AMiBA sensitivity}

I have adopted cosmological parameters $\Omega_{m} = 0.37$, 
$\Omega_{\Lambda} = 0.63$, $\Omega_{B}h^{2} = 0.02$ and $h = 0.7$ 
for computing CMB anisotropy spectra in order to
estimate the expected signal variance in AMiBA images.
A $C_{l}$ power spectrum decomposition of the SZE anisotropy expected
in this cosmological model was provided, for these calculations,
by Ue-Li Pen (private communication).

In sky images synthesized using the 0.3~m aperture array, the E-mode
CMB polarization signal is expected at appear with an rms of about 
5~mJy~beam$^{-1}$.  The 171-baseline array with these apertures may
image this linearly polarized signal, at 1-$\sigma$, in about 
18~hr integration per pointing.

An SZE cluster survey with the 171-baseline array made of 1.2-m apertures
would be confusion limited, in about 1~hr, because the thermal noise
contribution in this time would equal the T-mode confusion noise in the
synthetic images. Added in quadrature, the rms image noise in 1-hr 
integration time would be
2.0~mJy~beam$^{-1}$ and the rms anisotropy signal 
owing to cosmological clusters,
with a contribution about 0.5~mJy~beam$^{-1}$, would be buried in the
noise.  However, a simpleminded filtering by 
rejecting the 42 1.2-m baseline data would reject the
T-mode confusion and in images synthesized using the $(171-42)$ baseline
visibilities, the 0.3~mJy~beam$^{-1}$ SZE signal may be detectable
at 1-$\sigma$ in about 30~hr integration in any pointing.  It may be
noted here that the cosmological evolution in cluster gas properties
is poorly understood and predictions of the $C_{l}$ spectra expected
from cosmological clusters may have large uncertainties.  Consequently,
estimates of integration times required for detection of the SZE $C_{l}$s
are crude.

\section{Mosaicing}

Scanning the sky with interferometers has long been recognized as
a method for decomposing visibility measurements made with finite
apertures into equivalent visibilities which may be obtained by
dividing the apertures into smaller sections (Ekers \& Rots 1979,
Rao \& Velusamy 1984).  Interferometric measurements of CMB anisotropy,
which are made with single pointings, provide estimates of the CMB anisotropy
power spectrum integrated in $l$-space over a range (defined by a telescope
filter function) which corresponds to the auto-correlation function
of the aperture illumination.  A mosaic-mode observation of the CMB anisotropy
is  a method for decomposing the anisotropy measurements in 
$l$-space (White, Carlstrom, Dragovan \& Holzapfel 1999) and improving
the $l$-space resolution.

Short spacing interferometers have usually been
used in interferometric CMB
anisotropy measurements to give higher surface
brightness sensitivity.  However,
these suffer from an undesired response to the environment       
(brightness variations across the ground-sky interface and across
terrestrial objects) and are also susceptible to 
cross-talk (Subrahmanyan 2002).  The
unwanted response usually dominates the visibility due to the CMB anisotropy.
The suggested observing strategy with the AMiBA is drift scanning: the
co-mounted array is to be kept stationary with respect to the environment,
and the apertures are kept stationary with respect to each other, during
the data acquisition.  The unwanted environmental response 
and cross-talk is then to be filtered out of the data stream as a constant.

We believe that this observing strategy has its advantages as compared to
tracking arrays with independently mounted elements as well as tracking 
platforms with co-mounted apertures.
First, although tracking arrays
may differentiate between the sky signal and spurious coupling by
the difference in their fringe rates, the summed environmental coupling
and cross-talk would give an unwanted visibility contribution which
would inevitably have a slow time varying amplitude.  However, for
a drift-scanning array of co-mounted elements, 
the unwanted component would be a constant
in amplitude and phase and its subtraction is potentially exact.
Second, any baseline in a  tracking array
which has a small $v$-component, either because the baseline is north-south
oriented or because the observation is being made at a large hour angle,
would have difficulty separating the sky signal from the unwanted response.

\begin{figure}
\plotfiddle{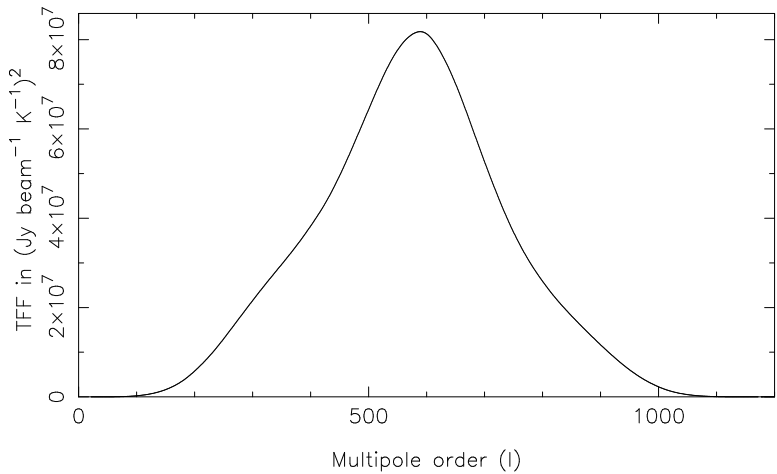}{6 truecm}{0.0}{60}{60}{-160}{-160}
\caption{Telescope filter function for a single pointing}
\end{figure}

Scanning with interferometers has added advantages: as noted above,
the time stream may be processed to improve the $l$-space resolution.
This creates the possibility that the spurious correlations and cross-talk,
which are expected to arise in the correlations between the fields
in those parts of the two apertures which are the nearest, may be
isolated.

Mosaic-mode observing, followed by a decomposition of the visibility
measurements in $u,v$-space, gives measurements in regions of $u,v$-space
which have poor sensitivity to sky anisotropy (because they represent
correlations between sections of the apertures which are poorly
illuminated by the feed).  These measurements provide useful `control samples' 
and may be used to estimate the instrument thermal noise from the
data.

\subsection{Sensitivity in Mosaic-mode observations}

The signal and noise are distributed differently across $l$-space or,
equivalently, $u,v$-space.  The decomposition of the data obtained 
in multiple sky pointings --- perhaps during a scanning of a 
sky region with the primary beam --- into data distributed in $l$-space
results in a separation of data with different signal-to-noise ratios.
The signal-to-noise ratios in the individual $l$-space filters
depends on the $C_{l}$ spectrum of the CMB anisotropies
and on the aperture illumination function.

As an illustration of the mosaic-mode observing, a particular case is 
considered here.
A two element interferometer is considered, which is made of two 0.3-m
apertures, and which are configured close-packed on a common
steerable platform to give a 0.3~m baseline.  This may represent the
interferometer being constructed as a prototype for the AMiBA.  Equipped
with the 95~GHz AMiBA receivers, the thermal rms noise in the 
real and imaginary parts of the complex Stokes parameter visibilities 
is expected to be 0.29~Jy for 1-hr integration time.

A measurement of the CMB anisotropy signal, in a single pointing and
with this single baseline, corresponds to a filtering of the CMB anisotropy
spectrum with a telescope filter function (TFF) given in Fig.~1. This
single filter defines the sensitivity, in $l$-space,  of 
the interferometer formed between the pair of 0.3~m apertures.
Assuming a `flat-band' CMB power with 
$l(l+1)C_{l}/(2 \pi) = 4 \times 10^{-12}$ ($(\Delta T/T)^{2}$; 
dimensionless), the expected 
signal is 33~mJy rms in the real and imaginary visibilities.
In an observation of a single pointing, 39~hr integration would be 
expected to yield one estimate of the anisotropy power --- averaged
in $l$-space over the range defined by the aperture extent --- with 
unity signal-to-noise ratio.

Consider next a mosaic-mode observation with the two-element
interferometer in which an RA strip of the sky is observed in a 
sequence of eleven pointings spaced 18.1~arcmin apart (the primary beam
FWHM is about 44~arcmin).  I assume that the platform is steered,
in RA, declination, and about a rotation axis, to keep the baseline
oriented parallel to the scan direction.   Assuming an integration
time of 39/11~hr in each pointing (and 39~hr in total),
each of the 11 complex visibilities would have a thermal rms noise of
0.15~Jy in the real and imaginary components.

The 11 complex visibilities {$X(n)$, $n$=$-5$, $-4$, ...., 4, 5} are 
discrete Fourier transformed:

$$Y(k) = \sum_{n=-5}^{n=+5} \lbrack X(n) 
e^{\lbrace i 2 \pi {{n k}\over{N}} \rbrace} \rbrack,$$

\noindent with $N=11$, to give eleven complex 
visibilities $Y(k)$, $k$=$-$5, $-$4, ..., 4, 5, 
distributed in $l$-space.  These eleven visibilities are measures
of the sky anisotropy signal as viewed through a bank of filters,
shown in Fig.~2, which span the $l$-space range covered by the
pair of apertures forming the interferometer.

\begin{figure}
\plotfiddle{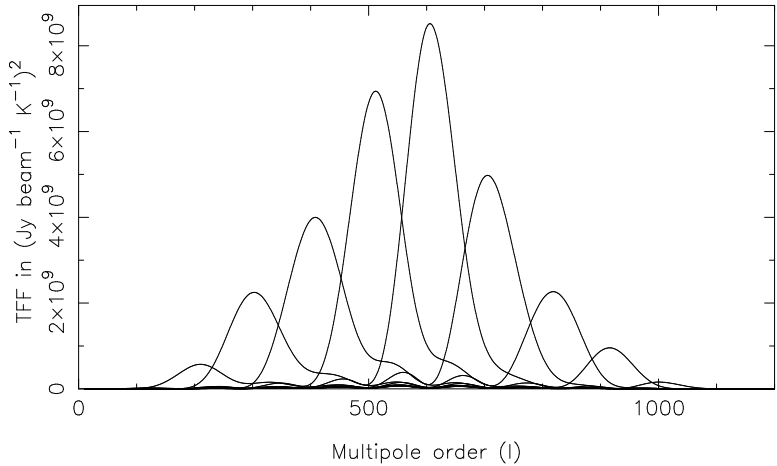}{6 truecm}{0.0}{60}{60}{-160}{-160}
\caption{Telescope filter functions for a 11-point mosaic}
\end{figure}

Adopting once again a `flat-band' CMB anisotropy spectrum, with
$l(l+1)C_{l}/(2 \pi) = 4 \times 10^{-12}$, the expected rms signal
amplitudes, and the signal-to-noise ratios, for the estimate of the
signal in each filter, for the eleven measurements $Y(k)$, 
are given in Table~1.

\begin{table}
\caption{Signals (S in mJy) and the signal to noise (S/N) expected
in the $l$-space filter bank}
\begin{tabular}{l r r r r r r r r r r r}
\tableline
Filter num. & $-$5 & $-$4 & $-$3 & $-$2 & $-$1 & 0 & 1 & 2 & 3 & 4 & 5 \\
\tableline
S & 36 & 80 & 131 & 152 & 170 & 169 & 125 & 82 & 52 & 31 & 27 \\
S/N & 0.1 & 0.2 & 0.4 & 0.4 & 0.5 & 0.5 
& 0.3 & 0.2 & 0.1 & 0.1 & 0.1 \\
\\
\tableline
\tableline
\end{tabular}
\end{table}

\noindent The rms thermal noise in each of these measurements, assuming
39/11~hr integration on each of the 
11 pointings, is $0.15 \times \sqrt{11} / \sqrt{2} = 0.36$~Jy.

In place of the eleven measurements $X(n)$, each 
with a signal-to-noise ratio of $1/\sqrt{11}$ and with independent 
noise, the mosaic-mode analysis which combines
the information from the multiple pointings simultaneously provides 
eleven estimates $Y(k)$ of the CMB anisotropy power spectrum
distributed in $l$-space with varying signal-to-noise ratios and with
independent noise.  

It may be noted that the central filters detect the CMB flat-band
power with signal-to-noise exceeding $1/\sqrt{11}$ where as the
edge channels respond poorly to sky anisotropy.  However, the
signal-to-noise in the detection of anisotropy in any of the $l$-space
filters generated via the mosaic-mode observation does not
exceed the signal-to-noise attainable if the total time were used
in a single pointing.  The sum of the squares of the
signal-to-noise ratios in the 11 filters, listed in Table~1, is unity,
and this is true for any plausible form for the
$C_{l}$ spectrum.  The mosaicing strategy evaluated here effectively 
redistributes the  observing time equally 
among the synthetic $l$-space filters
so that the sum of the squares of the signal-to-noise ratios over
all independent filters is invariant and is proportional to the total
observing time.  

\acknowledgments

The observing schemes discussed here evolved in discussions
between members of the AMiBA science and engineering teams.

\end{document}